\documentclass[letter]{aa} 

\usepackage{graphicx, comment}
\usepackage{commath}
\usepackage{txfonts}
\usepackage{subfig}
\usepackage{upgreek}
\usepackage[colorlinks=true]{hyperref}

\hypersetup{
     colorlinks   = true,
     citecolor    = blue
}

\usepackage{orcidlink}

\newcommand{\C}{3C\,84}

\newcommand{\smaxnum}{8.7\,\textrm{Jy/beam}}
\newcommand{\smaxnumnrao}{2.5\,\textrm{Jy/beam}}
\newcommand{\beam}{(0.17\times0.4)\,\textrm{mas}}
\newcommand{\beamnrao}{(0.20\times0.35)\,\textrm{mas}}
\newcommand{\rms}{0.4\,\textrm{mJy/beam}}
\newcommand{\rmsH}{0.3\,\textrm{mJy/beam}}
\newcommand{\rmsnrao}{0.4\,\textrm{mJy/beam}}
\newcommand{\rmsP}{0.8\,\textrm{mJy/beam}}
\newcommand{\rmsPH}{0.3\,\textrm{mJy/beam}}
\newcommand{\corrang}{93^\circ}
\newcommand{\angunc}{15^\circ}

\begin{document} 

    \title{Evidence of a toroidal magnetic field in the core of \C}

   \author{
   G.~F. Paraschos\inst{1}\orcidlink{0000-0001-6757-3098}, 
   L.~C. Debbrecht\inst{1}\orcidlink{0009-0003-8342-4561}, 
   J.~A. Kramer\inst{2,1}\orcidlink{0009-0003-3011-0454}, 
   E. Traianou\inst{3,1}\orcidlink{0000-0002-1209-6500}, 
   I. Liodakis\inst{4,5}\orcidlink{0000-0001-9200-4006},
   T.~P. Krichbaum\inst{1}\orcidlink{0000-0002-4892-9586},
   J.-Y. Kim\inst{6,1}\orcidlink{0000-0001-8229-7183}, 
   M. Janssen\inst{7,1}\orcidlink{0000-0001-8685-6544},  
   D.~G. Nair\inst{8}\orcidlink{0000-0001-5357-7805},
   T. Savolainen\inst{9,10,1}\orcidlink{0000-0001-6214-1085},
   E. Ros\inst{1}\orcidlink{0000-0001-9503-4892},
   U. Bach\inst{1}\orcidlink{0000-0002-7722-8412}, 
   J.~A. Hodgson\inst{11}\orcidlink{0000-0001-6094-9291},
   M. Lisakov\inst{1,12, 13}\orcidlink{0000-0001-6088-3819},
   N.~R. MacDonald\inst{14}\orcidlink{0000-0002-6684-8691},
   J.~A. Zensus\inst{1}\orcidlink{0000-0001-7470-3321}
          }

   \authorrunning{G.~F. Paraschos et al.}
   \institute{
              $^{1}$Max-Planck-Institut f\"ur Radioastronomie, Auf dem H\"ugel 69, D-53121 Bonn, Germany\\ 
              $^{}$\ \email{gfparaschos@mpifr-bonn.mpg.de}\\
              $^{2}$ Theoretical Division, Los Alamos National Laboratory, Los Alamos, NM 87545, USA\\
              $^{3}$ Instituto de Astrofísica de Andalucía-CSIC, Glorieta de la Astronomía, E-18008 Granada, Spain\\
              $^{4}$ NASA Marshall Space Flight Center, Huntsville, AL 35812, USA\\
              $^{5}$ Institute of Astrophysics, Foundation for Research and Technology - Hellas, Heraklion, GR7110, Greece\\
              $^{6}$ Department of Astronomy and Atmospheric Sciences, Kyungpook National University, Daegu 702-701, Republic of Korea\\
              $^{7}$ Department of Astrophysics, Institute for Mathematics, Astrophysics and Particle Physics (IMAPP), Radboud University, P.O. Box 9010, 6500 GL Nijmegen, The Netherlands\\
              $^{8}$Astronomy Department, Universidad de Concepción, Casilla 160-C, Concepción, Chile\\
              $^{9}$Aalto University Department of Electronics and Nanoengineering, PL 15500, 00076 Aalto, Finland\\
              $^{10}$Aalto University Metsähovi Radio Observatory, Metsähovintie 114, 02540 Kylmälä, Finland\\
              $^{11}$Dept. of Physics \& Astronomy, Sejong University, Guangjin-gu, Seoul 05006, Republic of Korea\\
              $^{12}$Lebedev Physical Institute of the Russian Academy of Sciences, Leninsky prospekt 53, 119991 Moscow, Russia\\
              $^{13}$Instituto de F\'{i}sica, Pontificia Universidad Cat\'{o}lica de Valpara\'{i}so, Casilla 4059, Valpara\'{i}so, Chile\\
              $^{14}$Department of Physics and Astronomy, The University of Mississippi, University, Mississippi 38677, USA
             }

   \date{Received -; accepted -}

 \abstract
   {
   The spatial scales of relativistic radio jets, probed by relativistic magneto-hydrodynamic (RMHD) jet launching simulations and by most very long baseline interferometry (VLBI) observations differ by an order of magnitude.
   Bridging the gap between these RMHD simulations and VLBI observations requires selecting nearby active galactic nuclei (AGN), the parsec-scale region of which can be resolved.
   The radio source \C\ is a nearby bright AGN fulfilling the necessary requirements: it is launching a powerful, relativistic jet powered by a central supermassive black hole, while also being very bright.
   Using 22\,GHz globe-spanning VLBI measurements of \C\ we studied its sub-parsec region in both total intensity and linear polarisation to explore the properties of this jet, with a linear resolution of $\sim0.1$\,parsec.
   We tested different simulation set-ups by altering the bulk Lorentz factor $\Gamma$ of the jet, as well as the magnetic field configuration (toroidal, poloidal, helical).
   We confirm the persistence of a limb brightened structure, which reaches deep into the sub-parsec region.
   The corresponding electric vector position angles (EVPAs) follow the bulk jet flow inside but tend to be orthogonal to it near the edges. 
   Our state-of-the-art RMHD simulations show that this geometry is consistent with a spine-sheath model, associated with a mildly relativistic flow and a toroidal magnetic field configuration.
   }


   \maketitle

\section{Introduction}

Making images of supermassive black hole \citep[SMBH;][]{EHT19a, EHT22a, EHT24} shadows and their associated relativistic jets \citep{Lu23} have significantly improved our understanding of how these jets are launched. 
Such a direct approach has only been possible for a select few radio sources; for all other sources, polarimetric observations are key for studying energy extraction from black holes.
A SMBH is thought to be capable of launching jets, both via its ergosphere \citep[][BZ model]{Blandford77} and via its accretion disc \citep[][BP model]{Blandford82}. 
A stratified combination \citep{Sol89, Laing96} of the two mechanisms can result in a spine-sheath type of flow \citep[see also][]{Hirotani24}.
This type of geometry has been revealed in a number of jets \citep[see e.g.][]{Attridge99, Gabuzda00, Pushkarev05, OSullivan09, Mertens16, Walker18, Janssen21, Lu23} and attempts have been made to model it \citep{Laing96, Murphy13,Laing14,Fuentes18}.

The relativistic plasma and magnetic fields in these AGN jets generate incoherent synchrotron emission across a range of wavelengths, from radio to optical, UV, and possibly X-rays, characterised by both linear and circular polarisation. 
Linear polarisation can reach values as high as 60-70\% of the total intensity flux, as noted by \citealt{Rybicki79}.
These polarisation characteristics provide insights into the jet’s physical properties, such as magnetic field intensity and configuration \citep{Tsunetoe20}, structural layout, particle concentration, and plasma makeup, as detailed by \cite{Wardle98}.

Images obtained in polarised emission are useful probes of jet physics because, in the non-relativistic case, the electric vector position angle (EVPA) is expected to align perpendicularly (in the optically thin regime) to the local magnetic field, aiding in understanding the magnetic field's structure within the source \citep[e.g.][]{2024A&A...682A.154T}. 
In the case of the spine-sheath model, the EVPAs are oriented along the spine and perpendicular to the sheath.
Relativistic magneto-hydrodynamic simulations \citep[see e.g.][]{Fuentes18, Kramer21} indicate that toroidal and helical magnetic field configurations are capable of producing such an EVPA configuration.
In this case the envelope of the magnetic field lines in the sheath corresponds to the observed poloidal component, whereas the loops correspond to the toroidal component.

Furthermore, using recent observations from the Imaging X-ray Polarimetry Explorer \citep[IXPE,][]{Weisskopf2022} \cite{Liodakis2022} have found that the electric vector polarisation angle of the millimetre-radio to X-ray emission is typically aligned with the jet axis. 
This was found to be true for the optically thin synchrotron emission in all sub-classes of blazars \cite[e.g.][]{Ehlert2023,Middei2023,Peirson2023}. 
The IXPE observations have also discovered a rapid rotation of the polarisation angle in X-rays without an optical or radio counterpart \citep{DiGesu23}. 
The lack of a counterpart at lower frequencies points to non-cospatial emission regions, which is consistent with energy-stratified emission from a shock-in-jet scenario \citep{Marscher85}.

At a luminosity distance of 78.9\,Mpc ($z=0.0176$, \citealt{Strauss92}) in the Perseus galaxy cluster, \C\ (NGC\,1275) is one of the nearest and brightest radio galaxies in the northern sky, and thus is a prime target to study jet formation.
Its parsec-scale jet is characterised by limb brightening \citep{Nagai14, Giovannini18}, reaching deep into the sub-milliarcsecond (sub-mas) and contributing to its complex morphology \citep{Hodgson18, Paraschos21, Punsly21, Oh22}.
Polarisation studies of the source have been carried out in the past \citep{Homan04, Taylor06, Plambeck14, Nagai17, Kim19, Paraschos24}, providing insights into the interaction of the jet with the ambient medium, as well as the accretion state of the central SMBH.

The radio source \C\ has been monitored at centimetre wavelengths for decades \citep[e.g.][]{Marr89, Vermeulen94, Walker94, Dhawan98, Walker00, Giovannini18, Britzen19, Paraschos22, Savolainen23} to study the parsec-scale jet structure\footnote{At greater distances from the central engine, on the decaparsec to kiloparsec scales, even more structure is detected \citep{Walker00, Fabian00}.}.
Three main areas of the approaching (southern) jet can be distinguished \citep[e.g.][]{Nagai14}; they are usually labelled as C1 (core), C2 (diffuse emission region south-west of C1), and C3 (a prominent, moving hotspot south of C1, associated with the restarted jet activity; \citealt{Nagai10}).
Polarisation has been detected in the compact sub-mas region and in the extended parsec-scale southern emission, although more prominently in the latter.
Specifically, the parsec-scale region has been reported to exhibit circular polarisation \citep{Homan04} and substantial Faraday rotation measures \citep{Taylor06}.
Newer measurements have shown that linear polarisation is also detected in C3 \citep{Nagai17} and in the ultimate vicinity of the central engine \citep{Paraschos24}.
Still, the polarisation signature in the core region of \C\ (C1) remains elusive, at distances of $\sim1\,\textrm{mas}$ from the central engine.

We thus utilised the excellent $(u, v)$-coverage and sensitivity offered by globe-spanning very long baseline interferometry (VLBI) to examine the core region of \C\ in both total and polarised intensity. 
We then compared our observations with state-of-the-art RMHD simulations to draw conclusions about the magnetic field configuration in that core region.

This letter is structured as follows. 
In Sect.~\ref{sec:Results} we briefly discuss our observations, data analysis, and results. 
In Sect.~\ref{sec:Disc} we compare the resulting image to RMHD simulations and draw conclusions about the magnetic field configuration.
Finally, in Sect.~\ref{sec:Conclusions} we present our conclusions.
Throughout the paper we assume a $\Lambda$ cold dark matter cosmology with $H_0 = 67.8\,\textrm{km s}^{-1}\,\textrm{Mpc}^{-1}$, $\Omega_\Lambda = 0.692$, and $\Omega_\textrm{M} = 0.308$ \citep{Planck16}.

\section{Data, analysis, and results}\label{sec:Results}

The radio source \C\ was observed in November 2021 with the European VLBI Network (EVN) in conjunction with other telescopes, to form a globe-spanning array, at both 22 and 43\,GHz, as part of a multi-wavelength effort to study its parsec-scale jet via direct imaging and polarisation.
In total, 22 antennas successfully observed the source (see Appendix~\ref{apps:Observations}).
Calibration and imaging of the data were performed using the \texttt{rPICARD} pipeline \citep{Janssen19}, the \texttt{polsolve} software \citep{MartiVidal21}, and the \texttt{difmap} software \citep{Shepherd94}.
Details of the post-correlation, calibration, and imaging procedures are discussed in Appendices~\ref{app:Calibration} and \ref{app:EVPA}.

\begin{figure*}
\centering
\includegraphics[scale=0.6]{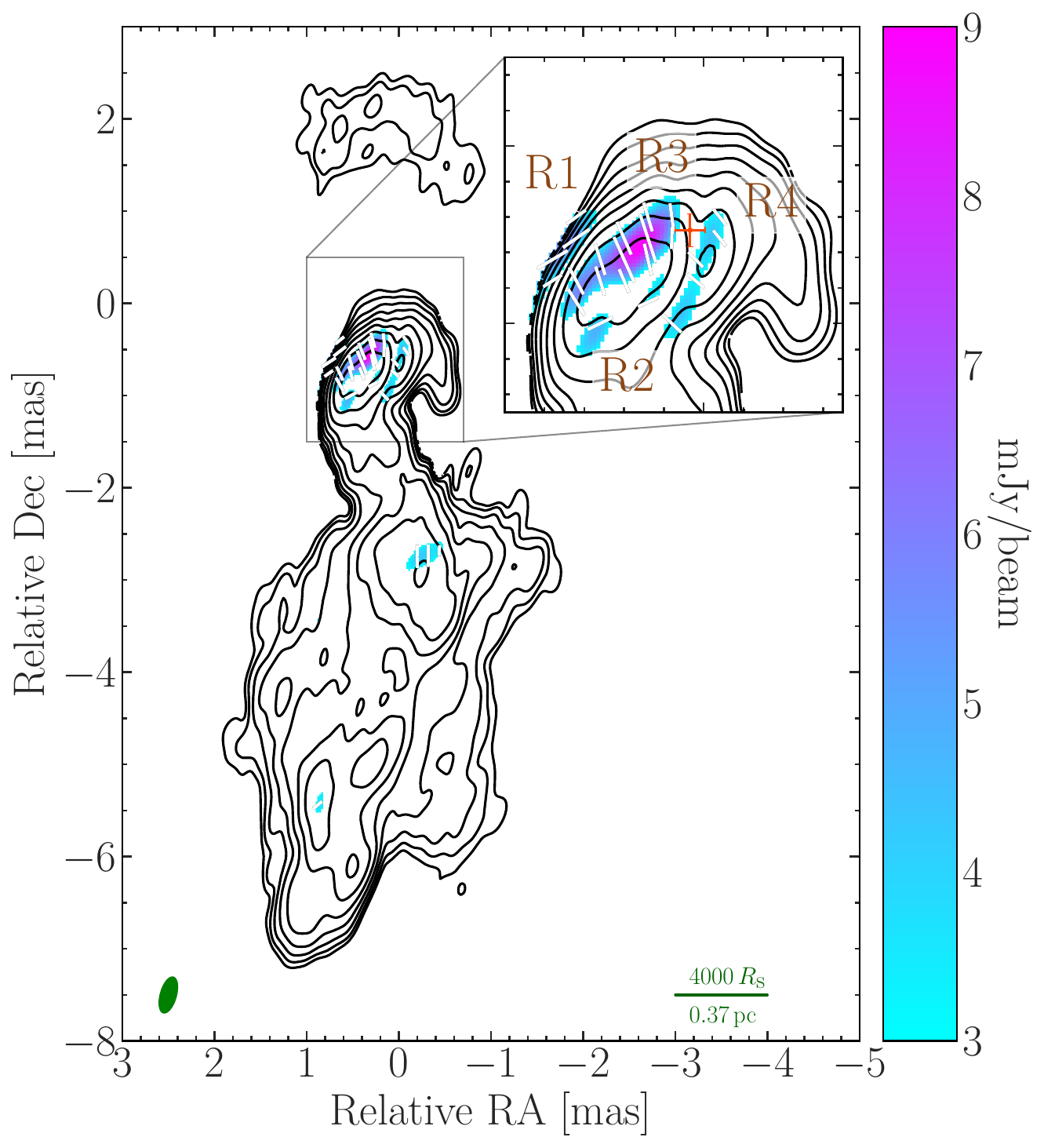}
  \caption{EVN image of Stokes $I$ (black contours), polarised intensity (colour scale), and EVPA (white bars) of \C\ at 22\,GHz.
  The contour levels are at 0.25, 0.5, 1, 2, 4, 8, 16, 32, and 64\% of the total intensity peak ($S_\textrm{max} = \smaxnum$).
  The lowest contour cutoff is at $5\,\upsigma_\textrm{I}$, with $\upsigma_\textrm{I} = \rms$.
  The convolving beam size is $\beam$ with a semi-major axis at a position angle of $15.8^\circ$ north to west (green ellipse in the bottom left corner); the green bar (bottom right) corresponds to a (projected) distance of $4000\,R_\textrm{S}$ (Schwarzschild radius).
  The cutoff is at $4\,\upsigma_\textrm{P}$, with $\upsigma_\textrm{P} = \rmsP$, and only the flux within the boundaries of the Stokes $I$ map is considered.
  The inset offers a zoomed-in view of the sub-mas region; the red cross corresponds to the position of the jet apex at 22\,GHz, as extrapolated in the core-shift analysis presented in \cite{Paraschos21}.
  For this we assumed that the brightest component in the core region of these observations can be identified with the brightest component in the 86\,GHz observations in that paper \citep[see also][]{Paraschos23}, shifted due to opacity effects.
  The denoted regions R$1-4$ are defined as R$1$: jet bend, R$2$: inner spine, R$3$: eastern sheath, and R$4$: western sheath. 
  }
     \label{fig:Pol}
\end{figure*}

Here we present the 22\,GHz image of \C, as observed in Stokes $I$ and linear polarisation (see Fig.~\ref{fig:Pol}).
Our observations constitute some of the highest resolution linearly polarised flux detections in AGN.
Similar observations in sensitivity and $(u,v)$-coverage were conducted as part the Global VLBI Alliance forum \citep[GVA;\footnote{\url{http://gvlbi.evlbi.org/}}][]{Park23} a year later, and exhibit very similar structure in total intensity.
That work also provides the most recent estimate of the bulk jet flow direction; the authors show that the jet propagates south-eastwards in the compact core region, before bending southwards further downstream.

We find evidence of linearly polarised flux in the sub-mas region, tracing two limbs, the EVPAs of which are perpendicular to the bulk jet flow (regions R$3$ and R$4$ in Fig.~\ref{fig:Pol}).
Specifically, region R$3$ is detected at a significance level exceeding $5\,\upsigma$ in polarisation, whereas region R$4$ corresponds to a $4\,\upsigma$ detection.
This points to the prevalence of magnetic field lines running parallel to the direction of the southern jet propagation.
Weak polarised flux is also detected between these two limbs, denoted as region R$2$.
Interestingly, the EVPAs seem transverse to those in the two limbs, indicating that the magnetic field there is perpendicular to the bulk jet flow and possibly indicative of a shock. 
A fourth region (R$1$) with polarised flux is also detected in the vicinity of where the jet curves southwards.
We discuss the implications of our findings and compare them to simulations in the next section.

\section{Discussion}\label{sec:Disc}

\subsection{Spine-sheath geometry}

Diverse mechanisms can give rise to a limb-brightened geometry, for example a velocity gradient in the jet caused by shear acceleration at its boundary, jet stratification, geometrical projection effects, as well as the direct influence of the SMBH powering the jet \citep[see e.g.][for a relevant review]{Blandford19a}.
In the latter case, this geometry, manifested in both Stokes $I$ (see Fig.~\ref{fig:StI}) and linear polarisation, has implications for the conditions of the central engine (i.e. the nature of the accretion flow, the spin of the BH, and the magnetic state) and for the magnetic field configuration downstream, in the sub-parsec region.
To investigate the sub-milliarcsecond jet region, \cite{Takahashi18} performed a simulation of a steady, axisymmetric force-free jet. 
They found that a limb-brightened symmetric structure may be associated with an advection-dominated accretion flow (ADAF) and a magnetically arrested disc (MAD) around a rapidly spinning SMBH.
Their findings were confirmed observationally, using the Event Horizon Telescope (EHT), to resolve the core region of \C\ at 230\,GHz \citep[see][]{Paraschos24}.

On the other hand, in the downstream region, helical and toroidal magnetic field lines have been associated with an increase in linear polarisation towards the jet edges \citep{Pushkarev05, Lyutikov05}.
Specifically, it is expected that toroidal magnetic field geometries give rise to symmetric Stokes $I$ and linear polarisation profiles, whereas helical geometries cause asymmetric profiles \citep{Gabuzda20, Kramer21}.
In addition to the intrinsic magnetic field carried by the jet, shear from velocity gradients in the flow might also contribute to the observed geometry \citep[e.g.][]{Laing80b}.
The EVPAs associated with these profiles are expected to be bimodal, that is, perpendicular to the bulk jet flow on the bright limbs and parallel to it on the inner ridge \citep[][see also \citealt{Kramer21} for RMHD simulations revealing the  EVPA orientation for different magnetic field geometries]{OSullivan09}.

\begin{figure}
\centering
\includegraphics[trim={0 0 0 0}, clip, width=0.49\textwidth]{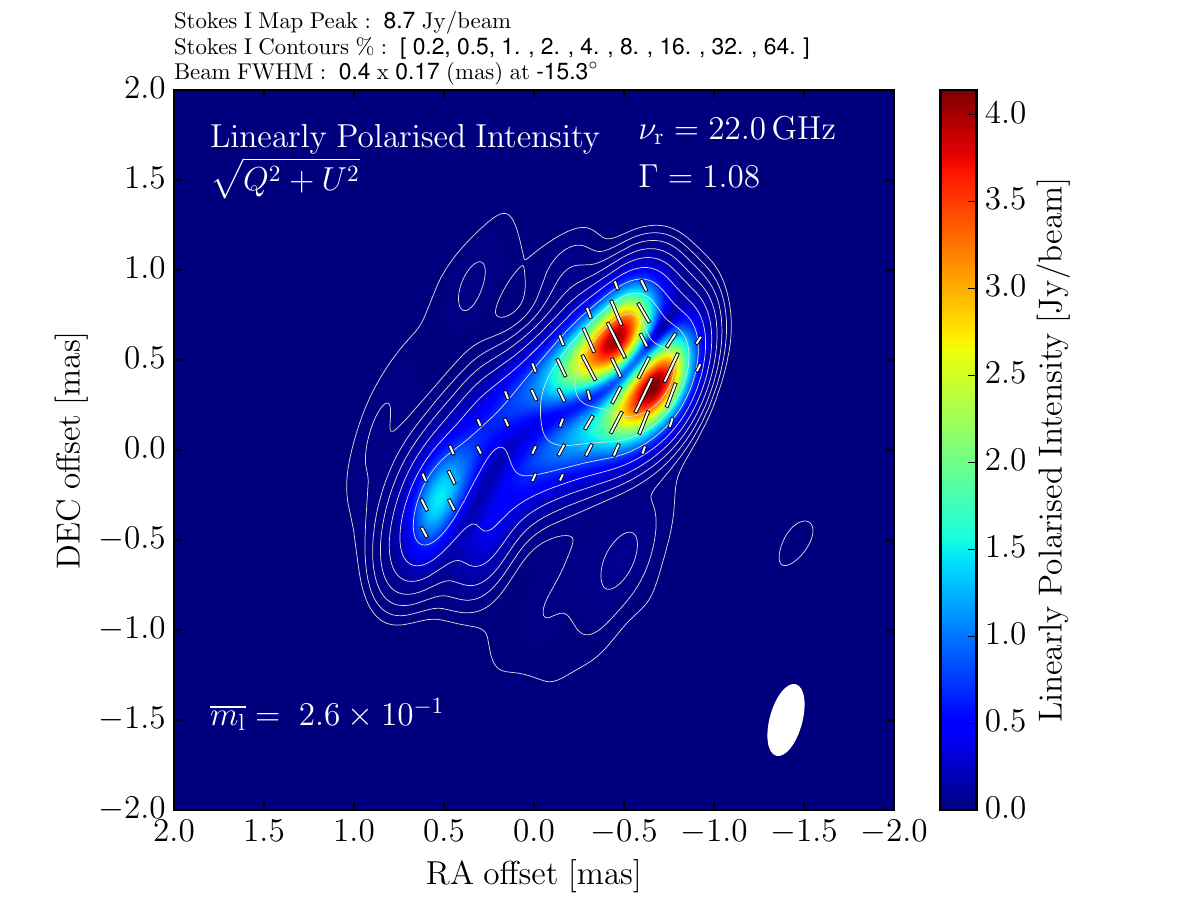} 
  \caption{
 Results from RMHD simulations. 
 A synthetic intensity map is shown of a ray-traced hybrid fluid-particle jet, moving from north-west to south-east, for a $\Gamma=1.08$.
 The colour bar sets the linearly polarised intensity calculated from the Stokes $Q$ and $U$ parameters. 
 The contours correspond to the total intensity flux density (Stokes $I$) and match those listed in Fig.~\ref{fig:Pol}.
 The image depicts the jet model as observed at a frequency of 22\,GHz. 
 The fractional linear polarisation ($\bar{m}_\textrm{l}$) level is indicated in the lower left corner.
 The convolving beam size (white ellipse, bottom right corner) equals the observational beam size.
 }
     \label{fig:Sim}
\end{figure}

Here we adapted these RMHD simulations to the parameters of \C. 
A more detailed description of the simulations set-up is described in Appendix~\ref{app:SimSetup}.
Specifically, we assume a luminosity distance of $D_\textrm{L} = 78.9\,\textrm{Mpc}$ and investigate the effects of different bulk Lorentz factors $\Gamma$ and magnetic field configurations \citep[toroidal, poloidal, helical; detailed descriptions and illustrations of the magnetic field vectors can be found in][]{Kramer21}.
For the fiducial simulation presented in Fig.~\ref{fig:Sim}, we chose a bulk Lorentz factor of $\Gamma=1.08$ \citep[the slow pattern speeds in the sub-mas region of \C, e.g.][suggest that $\Gamma\sim1$]{Krichbaum92, Dhawan98, Hodgson21, Weaver22, Paraschos22} and a toroidal magnetic field configuration. 
The figure shows linearly polarised intensity ($P = \sqrt{Q^2+U^2}$) overplotted with both total intensity contours and EVPAs ($\chi_\textrm{EVPA}=0.5\arctan(Q/U)$). 
The integrated level of fractional linear polarisation is shown in the lower left, that is $\bar{m_\mathrm{l}} = \sqrt{\bar{Q}^2+\bar{U}^2}/\bar{I}$ (barred quantities correspond to flux weighted averages).
In Appendix~\ref{app:SimParameters} we present alternative simulation results, based on these other parameters.
The uncertainty of the SMBH mass in \C\ (mass ranges between $M_\textrm{BH}=(3.2-20)\times10^8\,\textrm{M}_\odot$ have been reported in previous studies, e.g. \citealt{Park17, Giovannini18}) does not affect our simulations, given that general relativistic effects should be negligible at these distances from the central engine.

We find that the geometry of the EVPAs on the simulated double rail structure corresponds to a toroidal magnetic field geometry; this is also seen in our observations in Fig.~\ref{fig:Pol}.
The dominance of the toroidal magnetic field component beyond the recollimation region \citep{Laing14} is also a natural consequence of jet expansion, as the toroidal component's strength decreases more slowly than that of the poloidal component \citep{Burch79,Fuentes18}. 
This magnetic field geometry, as indicated by our observations, could imply a connection to magnetic jet launching, as is the case for the BZ model.
Our finding is in line with mounting evidence connecting the jet launching in \C\ to the BZ model \citep[e.g.][]{Takahashi18,Oh22,Paraschos23,Paraschos24}.

In light of the recent IXPE observations of blazars, our results also suggest that the spine dominates the radio to X-ray synchrotron emission, where electron populations are accelerated in shocks, hence the alignment of the polarisation vectors with the jet. 
The emission then becomes energy-stratified due to particle cooling, as the particles propagate downstream from the shock front. 
This would also suggest the X-ray polarisation angle rotation was possibly the result of a shock moving through the helical magnetic field \citep[see the discussion on the proposed model in][]{Kim24}.

\subsection{Comparison to observations}\label{ssec:Comparison}

These findings also agree well with our comparison of the observations to our hybrid fluid-particle jet model shown in Fig.~\ref{fig:Sim}, where we illustrate a ray-traced linearly polarised emission map of interpolated macro-particle attributes (a discussion of the used methodology is presented in the accompanying paper J.~A. Kramer et al. in prep.). 
Specifically, we could accurately identify three of the four regions in our simulation. 
The 45$^\circ$ EVPA pattern on the south-western limb of our simulation is explained by the rotational viewing angle of about 35.5$^\circ$. 
If the jet were viewed fully edge-on \cite[see e.g.][]{Kramer21} the EVPA pattern would be perpendicular to the jet direction. 
However, due to the rotation of the viewing angle, they appear as observed in region R$4$ of Fig.~\ref{fig:Pol}.
The north-eastern limb of our observations (region R$3$) is a feature that is also replicated with our simulations. 
In this case, it is predicted that the EVPAs should be perpendicular to the propagation direction, which is exactly what is observed with our VLBI data.
On the spine (region R$2$), the EVPAs are parallel to the jet propagation direction of the core region \citep{Park23}, as predicted in a fully toroidal simulation \cite[visible in Fig.~\ref{fig:Sim} and in][]{Kramer21}. 
The changing viewing angle might also explain the difference in total intensity observed between the two limbs (i.e. due to Doppler boosting).
On the other hand, if this difference is an intrinsic feature of the \C\ jet, a highly twisted magnetic field (i.e. not purely toroidal) could still reconcile the observations with our simulations.

Region R$1$ is not present in our simulations; however, the polarisation enhancement there could be connected to the jet colliding with the interstellar medium and abruptly turning southwards scenario that was proposed in \cite{Park23}.
Through a kinematic and spectral study the authors showed that the jet could be changing its direction of propagation in that area.
Specifically, they found that in that area the spectral index flattens out, hinting at the presence of a working surface of the radio jet.
In that case, the magnetic field would be compressed there, resulting in an increase in its strength, manifested in a linear polarisation signal.
Nevertheless, we note that our detection of the linearly polarised signal is marginal.
This might indicate that, alternatively, the bend of the jet is intrinsic and the spectral inversion seen in \cite{Park23} is due to opacity effects of the jet bend geometry seen at an oblique angle.
Variability might also play a role, as newly ejected jet components might have altered the core region morphology.
Our follow-up high-sensitivity EVN observations might be able to provide a more definitive conclusion.

\subsection{Counter-jet depolarisation}

Parallel to the signatures of polarised light detected in the core region of \C, it is also of interest to note the absence of such signatures in the counter-jet region, as seen in Fig.~\ref{fig:Pol}.
Here we assume the counter-jet to correspond to the diffuse emission seen north of the core region.
Weaker polarisation signals in the receding jet side have already been reported \citep[e.g.][]{Laing88} and Faraday depolarisation due to the presence of free-free absorbing circumnuclear gas \citep[see discussion of the ambient medium around \C\ in e.g.][]{Walker00,Fujita17,Wajima20,Kino18, Kino21,Savolainen23,Park23} as one possible explanation.
Using a noise estimated upper limit of $S_\textrm{P}^\textrm{cj}= 1.6\,\textrm{mJy/beam}$ for the linearly polarised flux of the counter-jet, we were able to compute the jet-to-counter-jet ratio of polarised light to be $\Uppi_\textrm{P}\sim5$ (under the assumption that the approaching and receding jet sides have intrinsically the same brightness).
This corresponds to an optically thick medium, characterised by an optical depth of $\tau_\textrm{ff}=-\ln\left(\Uppi_\textrm{P} \right)\sim1.6$ \citep{Rybicki79}, which obscures the counter-jet.
For comparison, we note that \cite{Wajima20} measured higher values at 43\,GHz and 86\,GHz, using the Korean VLBI Network (KVN), but they were computing the total intensity jet-to-counter-jet ratio.
Observations analysing the atomic and molecular gas content closing in on the central engine of \C, similar to those performed by \cite{Nagai19}, \cite{Morganti23}, and \cite{Oosterloo24} will shed more light on the nature of the obscuring medium.

\section{Conclusions} \label{sec:Conclusions}

In this work we compared high-fidelity high-resolution observations of the core region of \C\ in total intensity and linear polarisation to RMHD simulations.
Our findings are summarised as follows:
\begin{itemize}
    \item The sub-mas region of \C\ exhibits a limb-brightened structure both in total intensity and linear polarisation.
    \item Linearly polarised flux is also detected in the region between these two limbs; the associated EVPAs exhibit a bimodal structure.
    They follow the bulk jet flow between the limbs, but are perpendicularly oriented to it at the two limbs.
    \item Comparing our observations to RMHD simulations reveals that the jet flow is consistent with a spine-sheath geometry, associated with a mildly relativistic flow ($\Gamma\sim1$) and a toroidal magnetic field.
    \item A linearly polarised signal is detected at the site of a possible jet bend, implying an increase in the magnetic field strength there, due to the collimation of the magnetic field lines.
\end{itemize}

Overall, our study presents an exploration of how state-of-the-art RMHD simulations can be utilised to gain new insights into jet launching in nearby AGN.
In future studies, facilitated by the ever-increasing sensitivity of globe-spanning VLBI arrays, such as the EVN array, additional radio sources will be scrutinised in more detail.
In the same vein, our upcoming, multi-frequency follow-up observations of \C\ will shed more light on its core region by also utilising spectral information.

\begin{acknowledgements}
      We would like to thank the anonymous referee for their constructive comments, which improved our work.
      G. F. P. wishes to thank Dr. Junghwan Oh for his valuable contribution to the preparation of the proposal leading to the observations presented here, Ms. Hui-Hsuan Chung for helping with the facilitation of the observations at the Effelsberg 100m telescope, and Dr. Daewon Kim for his useful comments.
      This research is supported by the European Research Council advanced grant “M2FINDERS - Mapping Magnetic Fields with INterferometry Down to Event hoRizon Scales” (Grant No. 101018682). 
      J. A. K. is supported for her research by a NASA/ATP project. The LA-UR number is LA-UR-24-22727.
      The MPCDF high-performing cluster Raven was used for the simulations.
      J.-Y. K. is supported for this research by the National Research Foundation of Korea (NRF) grant funded by the Korean government (Ministry of Science and ICT; grant no. 2022R1C1C1005255).
      I. L. was supported by the NASA Postdoctoral Program at the Marshall Space Flight Center, administered by Oak Ridge Associated Universities under contract with NASA. 
      I. L. was funded by the European Union ERC-2022-STG - BOOTES - 101076343. 
      Views and opinions expressed are however those of the author(s) only and do not necessarily reflect those of the European Union or the European Research Council Executive Agency. Neither the European Union nor the granting authority can be held responsible for them. 
      D. G. N. acknowledges funding from Conicyt through Fondecyt Postdoctorando (project code 3220195).

      Partly based on observations with the 100-m telescope of the MPIfR (Max-Planck-Institut für Radioastronomie) at Effelsberg.

      The research leading to these results has received funding from the European Union’s Horizon 2020 research and innovation program under grant agreement No 101004719 [Opticon RadioNet Pilot ORP].

      The European VLBI Network is a joint facility of independent European, African, Asian, and North American radio astronomy institutes. 
      Scientific results from data presented in this publication are derived from the following EVN project code: GP058. 
      The VLBA is an instrument of the National Radio Astronomy Observatory. 
      The National Radio Astronomy Observatory is a facility of the National Science Foundation operated by Associated Universities, Inc. 

      The data were correlated at the correlator of JIVE in Dwingeloo, the Netherlands. 
      This research has made use of the NASA/IPAC Extragalactic Database (NED), which is operated by the Jet Propulsion Laboratory, California Institute of Technology, under contract with the National Aeronautics and Space Administration. 
      This research has also made use of NASA's Astrophysics Data System Bibliographic Services. 
      
      Finally, this research made use of the following python packages: {\it numpy} \citep{Harris20}, {\it scipy} \citep{2020SciPy-NMeth}, {\it matplotlib} \citep{Hunter07}, {\it astropy} \citep{2013A&A...558A..33A, 2018AJ....156..123A} and {\it Uncertainties: a Python package for calculations with uncertainties.
      }
\end{acknowledgements}

\bibliographystyle{aa} 
\bibliography{sources}

\begin{appendix}
\section{Observations}\label{app:AppData} 

\subsection{EVN 22 GHz data} \label{apps:Observations}

\begin{figure}
\centering
\includegraphics[scale=0.5]{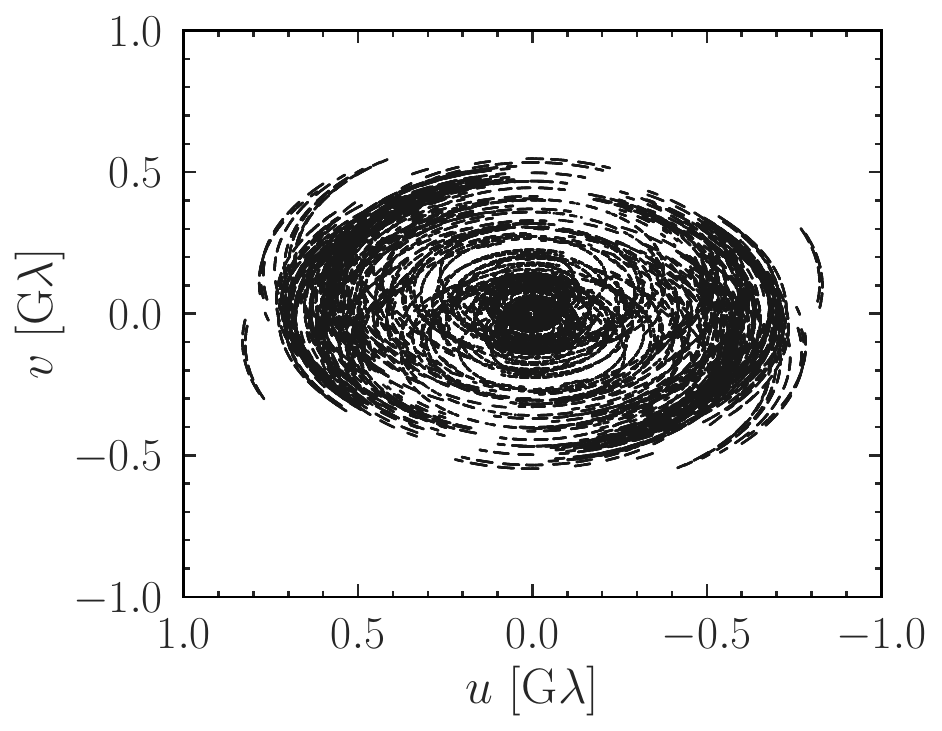}
  \caption{
  Illustration of the globe-spanning VLBI $(u, v)$-coverage of the target source \C.
  The density of the $(u, v)$-coverage increases the image fidelity.
  }
     \label{fig:uv}
\end{figure}

The radio source \C, along with the calibrator sources NRAO\,150 and J0102+58, was observed with the EVN, augmented by additional antennas forming a globe-spanning array, at 22\,GHz in dual circular polarisation in November 2021. 
The sources were observed by the following antennas: Effelsberg (EB), Jodrell Bank (JB), Onsala (ON), Mets\"ahovi (MH), Medicina (MC), Yebes (YS), Brewster (BR), Fort Davis (FD), Kitt Peak (KP), Los Alamos (LA), Mauna Kea (MK), North Liberty (NL), Owens Valley (OV), Pie Town (PT), Hancock (HN), Saint Croix (SC), KVN Tamna (KT), KVN Ulsan (KU), KVN Yonsei (KY), Svetloe (SV), Urumqi (UR), and Zelenchukskaya (ZC) in full track mode for a total of 26\,h. 
The data were recorded at an aggregate bitrate of 2\,Gbps with two-bit digitisation.
The resulting $(u, v)$-coverage for \C\ is shown in Fig.~\ref{fig:uv}.
The data were then channelised in eight intermediate frequencies (IFs) per polarisation, with an IF bandwidth of 32\,MHz; the data correlation was performed in zoom band mode resulting in a total recording bandwidth of $16 \times 32$\,MHz.
Finally, the data were correlated using the SFXC correlator \citep{Keimpema15} at Joint Institute for Very Long Baseline Interferometry European Research Infrastructure Consortium (JIVE).

\subsection{Calibration and imaging} \label{app:Calibration}

The next step in the data evaluation involved the post-correlation and calibration steps.
We used the \texttt{rPICARD} pipeline \citep{Janssen19} to perform the standard VLBI fringe-fitting and calibration procedures.
In short, we calibrated the phases by removing the delay and phase offsets between the IFs and then applied a global fringe fitting algorithm over the full bandwidth.
The visibility amplitudes were calibrated using the measured system noise temperature; the gain curve was also applied for each station.
For the stations that do not take atmospheric opacity effects into account, we applied the necessary corrections with the inferred receiver temperature and opacity along the line of sight.
Since we were interested in the polarised emission, we also removed the cross-hand phase and delay offsets.
More details about the VLBI calibration can be found, for example, in \cite{MartiVidal12}, \cite{Hada16}, and \cite{Kim19}.

The data were exported and then frequency-averaged and time-averaged (coherently, in 30-second bins) to be imaged using the \texttt{difmap} software \citep{Shepherd94}.
Outliers in the visibility domain were flagged.
A uniform weighting scheme was used in the images presented here, with the goal of sufficiently resolving the fine linearly polarised structures in the core.
Naturally weighted images were also produced to search for more extended linearly polarised signal in the outer regions; however, their S/N was not sufficient for a robust detection and we therefore refrain from presenting them here.
We then imaged the sources in total intensity, using the standard, iterative procedure of \texttt{CLEAN}ing and self-calibration in phase and amplitude.
Finally, the total flux was scaled up to $f_\textrm{comp}\sim90\%$ of the total single dish flux measurements, taken with EF before, during, and after the observations.
The compactness factor $f_\textrm{comp}$ was estimated based on a comparison between the 15\,GHz variability light curve \citep{Paraschos23} and the corresponding VLBI flux of \C\ \citep[][compare also with the MOJAVE\footnote{\url{https://www.cv.nrao.edu/MOJAVE/}} database \citealt{Lister18}]{Paraschos22}.
The uncertainty in the averaged single-dish measurement is of the order of 0.5\%.

\begin{figure}
\centering
\includegraphics[scale=0.3]{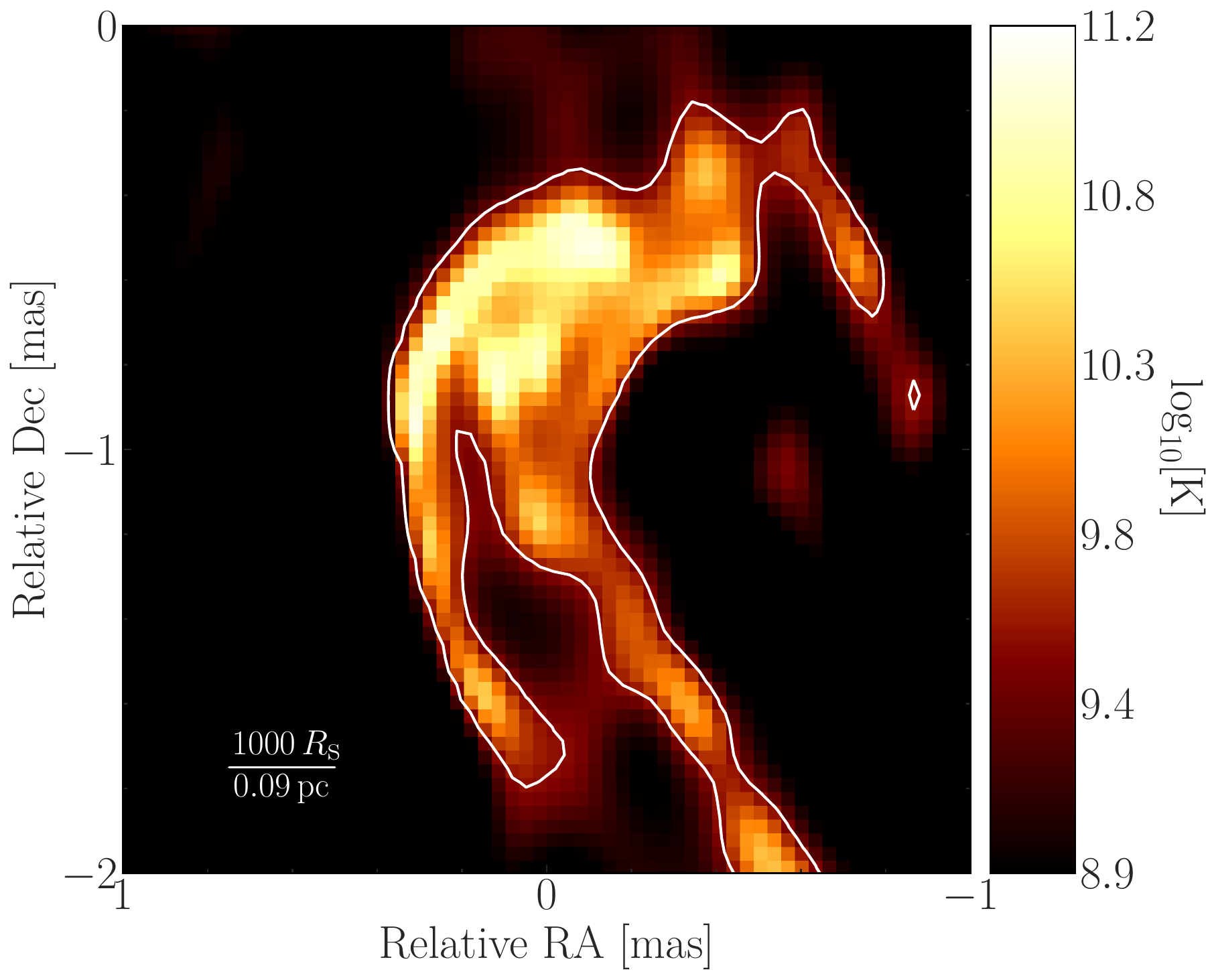}
  \caption{
  RML imaging result.
  A zoomed-in image of the core region of \C\ in Stokes $I$ is shown.
  The white contour denotes the 2\% flux level of the total intensity peak (of 0.06\,Jy/pixel).
  Here we utilised forward-modelling with \texttt{ehtim} to leverage the ability of RML methods to reconstruct continuous filaments faithfully.
  The two limbs reaching into the compact core region are clearly detected.
  The colour scale indicates the brightness temperature.
  }
     \label{fig:StI}
\end{figure}

In Fig.~\ref{fig:StI} we show a zoomed-in version of the limb brightening of \C\ in Stokes $I$. 
In order to better examine this high $S/N$ region, we reconstructed the image with the software package \texttt{ehtim} \citep{Chael_2016, 2018ApJ...857...23C}.
Among the forward modelling approaches, the regularised maximum likelihood (RML) method is an image synthesis family that require maximising the likelihood of a set of visibility data, given a set of predicted model visibility values and regularisers. 
RML imaging differs from traditional methods like \texttt{CLEAN} by offering a non-parametric approach to reconstructing images directly from the data, achieving better angular resolution and image fidelity \citep[e.g.][]{EHT19d,EHT22c, Fuentes23}.

For the first imaging iteration, we used a prior image of a Gaussian model with a size of 250\,$\upmu$as of full width at half maximum, across a 6.5\,mas field of view, and mapped on a $200\times200$ pixel grid. 
We incorporated closure phases and logarithmic closure amplitudes into the imaging process, accounting for non-closing systematic errors of 1\%. 
Having achieved convergence, we proceeded with self-calibration. 
In the second imaging run, we included visibility phases and amplitudes in the process. 
The third and last imaging round we included both closure quantities and complex visibilities. 
To ensure the correct choice of relative weights for the regularisation terms, we conducted a parameter survey, testing different values for the relative entropy parameter (MEM), total variation (tv), total squared variation (tv2), and the $\ell$1 norm \citep[see also][]{Savolainen23}. 
The final combination of values was chosen based on the criterion that provided the best fit to the data, producing the highest fidelity image at relative MEM: 0.1, tv: 0.3, tv2: 0, and $\ell$1: 0.1.

\subsection{Polarisation} \label{app:EVPA}

We utilised the \texttt{polsolve} software \cite{MartiVidal21} to compute the antenna instrumental polarisation solutions (D-terms).
Using all three sources (\C, NRAO\,150, and J0102+58) yielded the most robust set of D-terms, which were then vector averaged in order to obtain the nominal calibrated full Stokes images of the sources.  

\begin{figure}
\centering
\includegraphics[scale=0.4]{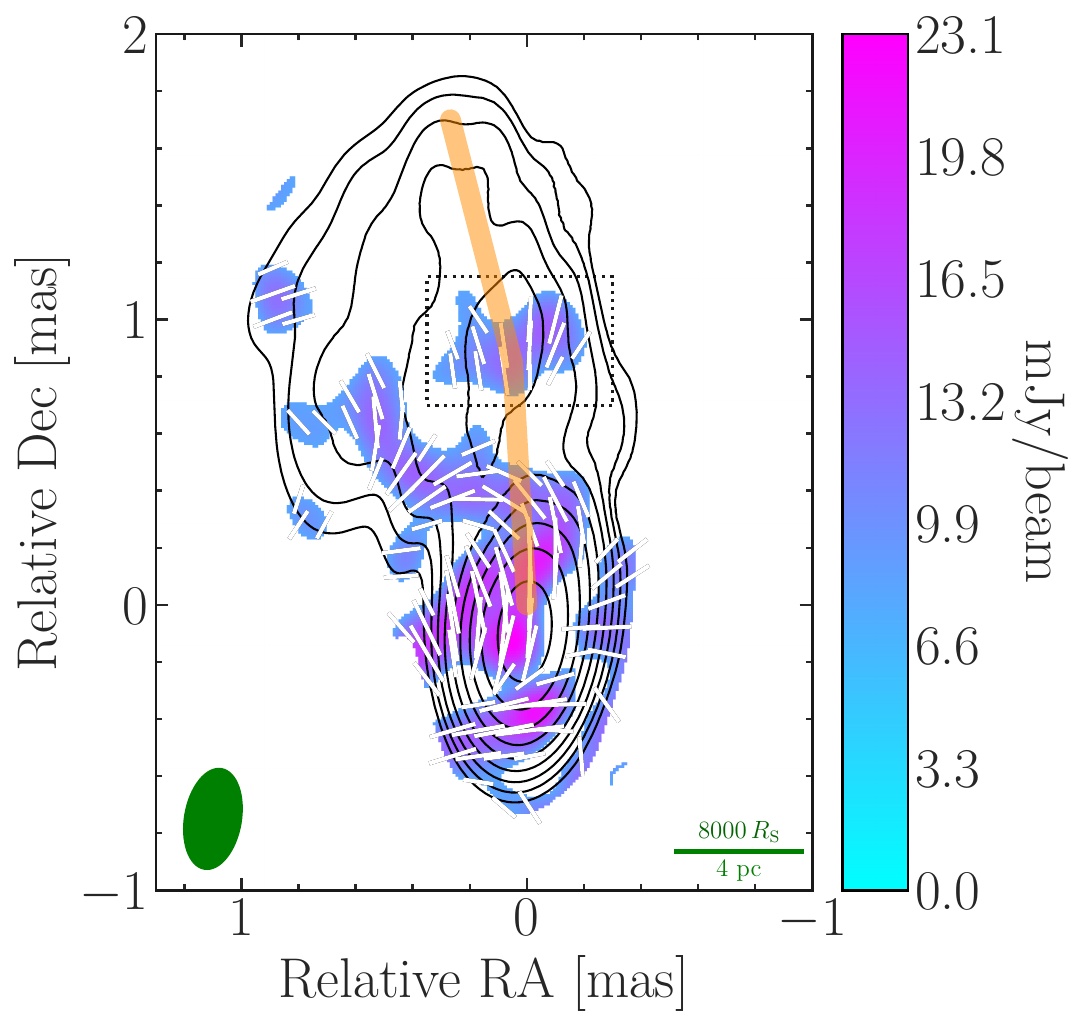}
  \caption{
  Stokes $I$ (black contours) and polarised intensity (colour scale) image of NRAO\,150 at 22\,GHz. 
  The contour levels at 0.25, 0.5, 1, 2, 4, 8, 16, 32, and 64\% of the total intensity peak ($S_\textrm{max} = \smaxnumnrao$). 
  The lowest contour cutoff is at $6\,\upsigma_\textrm{I}$, with $\upsigma_\textrm{I} = \rmsnrao$. 
  The green ellipse in the bottom left corner denotes the convolving beam size of $\beamnrao$ with a semi-major axis at a position angle of $10.9^\circ$ north to east.
  The green bar (bottom right) corresponds to a (projected) distance of $8000\,R_\textrm{S}$ (4\,pc). 
  The polarised intensity cutoff is at $6\,\upsigma_\textrm{P}$. 
  The dotted box corresponds to the optically thin, downstream jet region used to align the EVPAs to the bulk jet flow.
  The white bars show the orientation of the EVPAs with the calculated correction of $\Delta\upchi_\textrm{turn} \sim \corrang$. 
  The orange line denotes the ridge line.
  }
     \label{fig:nrao150}
\end{figure}

Determining the absolute positions of the EVPAs requires prior knowledge of quasi-simultaneous single-dish polarised flux measurements or an estimate of how the D-terms change with time.
At 22\,GHz no such measurements were available to us.
We therefore assumed that the EVPAs in the downstream jet area of the calibrator NRAO\,150 are intrinsically aligned to the bulk jet flow, as traced by the ridge line\footnote{See \cite{Paraschos22} for details about the ridge line determination.} \citep[see e.g.][]{Pushkarev05, Hada16}.
Figure~\ref{fig:nrao150} displays NRAO\,150 in Stokes $I$ and polarised flux density.
As shown in \cite{Molina14} for past observations at 22\,GHz, we note that some EVPA orientation variability with time is present in the core region.
In the downstream optically thin region, however, the detections are marginal and only in some epochs, so they do not allow an estimate of their temporal variability.
The calculated correction is $\Delta\upchi_\textrm{turn} \sim \corrang$.
We also note with interest the filamentary structure in linear polarisation present in this source (see Fig.~\ref{fig:nrao150}), which we will investigate in an upcoming work (Debbrecht et al. in prep.).

\begin{figure}
\centering
\includegraphics[scale=0.4]{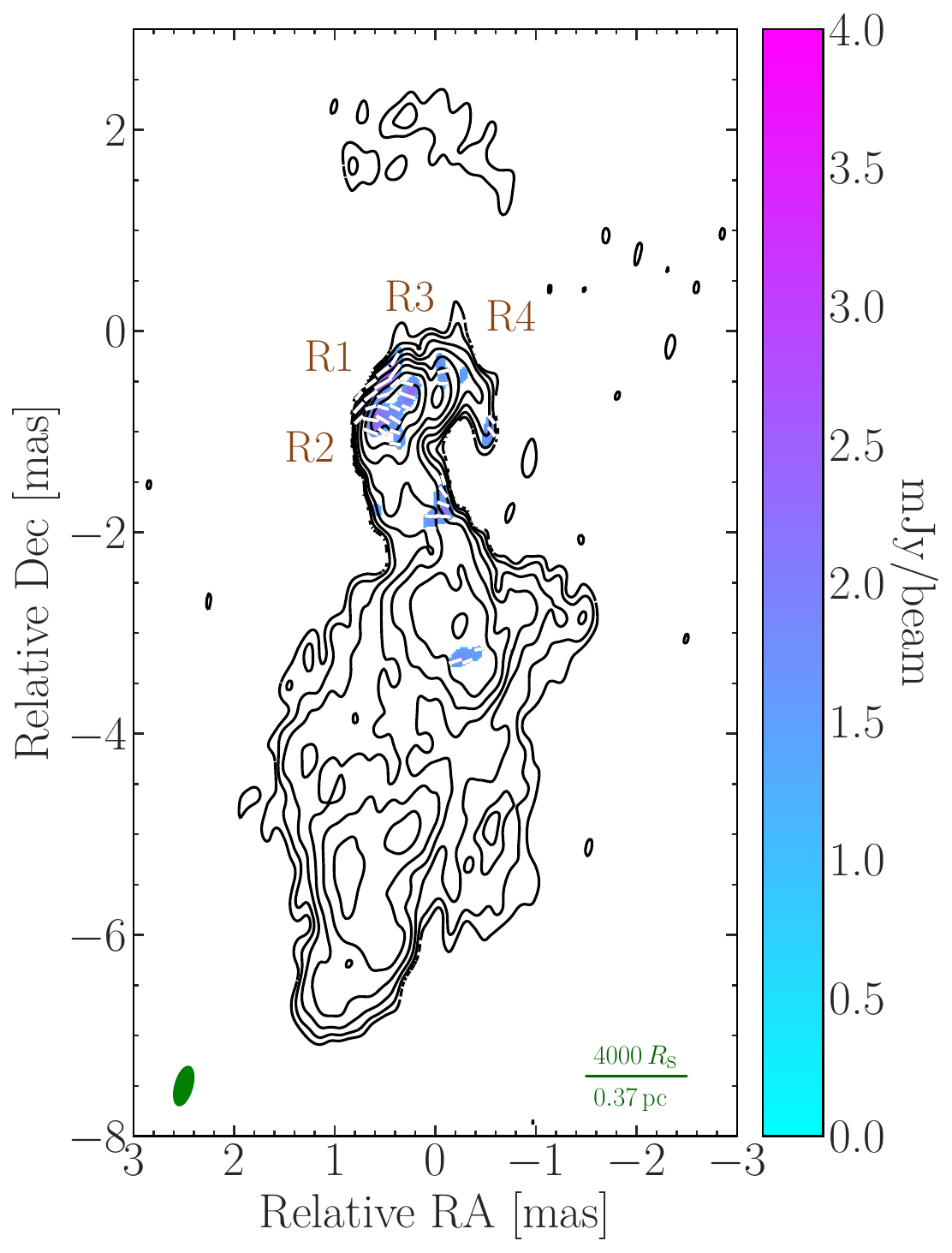}
  \caption{
  Alternative EVN image of Stokes $I$ (black contours), polarised intensity (colour scale), and EVPA (white bars) of \C\ at 22\,GHz, produced with the \texttt{Dog-Hit} pipeline.
  The plot set-up is the same as in Fig.~\ref{fig:Pol}; here $\upsigma_\textrm{I} = \rmsH$ and $\upsigma_\textrm{P} = \rmsPH$.
  Regions R$1$ and R$4$ are clearly discernible, whereas regions R$2$ and R$3$ are blended together.
  }
     \label{fig:PolH}
\end{figure}

Since inaccurately determined D-terms could affect the revealed spine-sheath polarisation pattern, we sought to determine them in an alternative way.
We used a pipeline based on the \texttt{DoG-HiT} software \citep{Mueller22}; this approach has the added benefit of also quantifying the uncertainty in the absolute position of the EVPAs.
The result is displayed in Fig.~\ref{fig:PolH}.
The EVPA patterns are in agreement with the fiducial image reconstruction shown in Fig.~\ref{fig:Pol}.
Region R$1$ of the jet bend is recovered, as is the jet sheath (regions R$3$ and R$4$).
Region R$2$, which in our interpretation corresponds to the spine, is blended together with R$3$ here.
A comparison between the EVPAs in the two implementations yields a total uncertainty of $\Delta\upchi_\textrm{abs} = \angunc$, while the average difference in corrections from the D-terms is of the order of $4\%$.
Our approach discussed above is based on a number of assumptions, yielding uncertainties in the presented absolute EVPA orientation.  
We note, however, that the EVPAs in the sheath and spine are at right angles, regardless of the correction factor applied to determine their absolute orientation.

\section{Simulations} \label{app:Simulations}

\subsection{Set-up} \label{app:SimSetup}

In the PLUTO Code \citep{Mignone07} the framework includes the equations of special RMHD, where a set of equations represents the vector of conservative variables and their associated fluxes: 
\begin{equation}
    \partial_t U^k + \partial_i T^{ik} = 0.
\end{equation}
These simulations operate with dimensionless grid units, requiring the conversion of the RMHD jet flow's thermal properties into physical units during post-processing. 
Using dimensionless quantities helps avoid extremely small or large numbers in the simulation. 
Nonetheless, to achieve an accurate representation, physical scaling to include specific scales of length, time, and energy is essential~\citep[see e.g.][]{Mignone07,Kramer21}.
RMHD simulations struggle to capture the kinetic-scale physics necessary for producing the non-thermal electron distributions behind observed synchrotron emissions. 
However, our recent jet simulations (see Fig.~\ref{fig:Sim} and J.~A. Kramer et al. in prep. for details), suitable for parsec-scale jets, incorporate microphysics to automatically create power-law electron distributions.
\cite{Vaidya} developed a novel hybrid framework for particle transport within the PLUTO code\footnote{For details refer to \url{http://plutocode.ph.unito.it}.}~\citep{Mignone07} that enhances the understanding of high-energy non-thermal emissions using Lagrangian particles from 3D RMHD simulations. This approach leverages the simulation's magnetic field to assess radiative losses, including synchrotron cooling, and integrates micro-physical aspects of spectral evolution based on fluid dynamics. The inclusion of Lagrangian particles in these simulations has proven instrumental in revealing detailed fluid characteristics such as velocity, density, and magnetic field strength, enabling an in-depth analysis of plasma transport, magnetic field impacts, and astrophysical phenomena like particle acceleration and shock dynamics. This methodology provides a more nuanced view of complex processes in astrophysical jets, aligning closely with observational data \citep{Joana}.

The robust ray-tracing code RADMC-3D, developed by \cite{Dullemond} for astrophysical radiative transfer issues, calculates the Stokes parameters $I$, $Q$, $U$, $V$ for total, linear, and circular polarisation. It handles synchrotron absorption, emissivity, Faraday conversion, and rotation \citep{MacDonald18}. 
The output, influenced by optical depth, normalisation constant, power-law distribution's low-energy cutoff, magnetic field strength, and orientation, offers a detailed view of polarised synchrotron radiation. 
Unlike \cite{Kramer21}, we used the enhanced Stokes module in RADMC-3D \citep{MacDonald21}, introducing a full rotational perspective to our 3D solution and adapting it to account for variable spectral indices. 
Finally, we performed a similar procedure for scaling the simulated flux densities to the observed ones, as discussed in Appendix~\ref{app:Calibration}.
We note that the simulated fractional linear polarisation is higher than that of the observations because in our simulations depolarising agents, such as foreground Faraday screens and free-free absorbing media, were not taken into consideration.

\subsection{Parameter comparison} \label{app:SimParameters}

\begin{figure*}
\centering
\includegraphics[trim={0 0 0 0}, clip, scale=0.45]{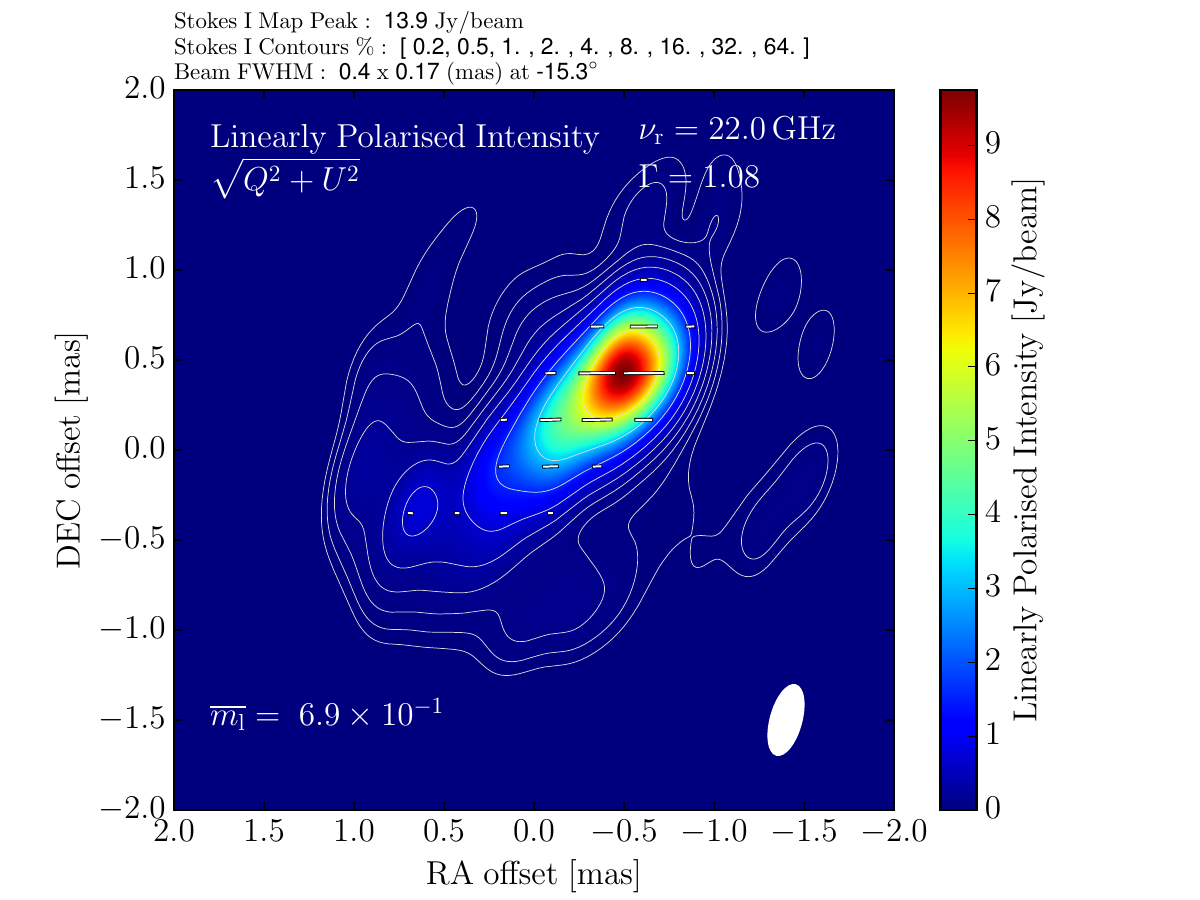} 
\includegraphics[trim={0 0 0 0}, clip, scale=0.45]{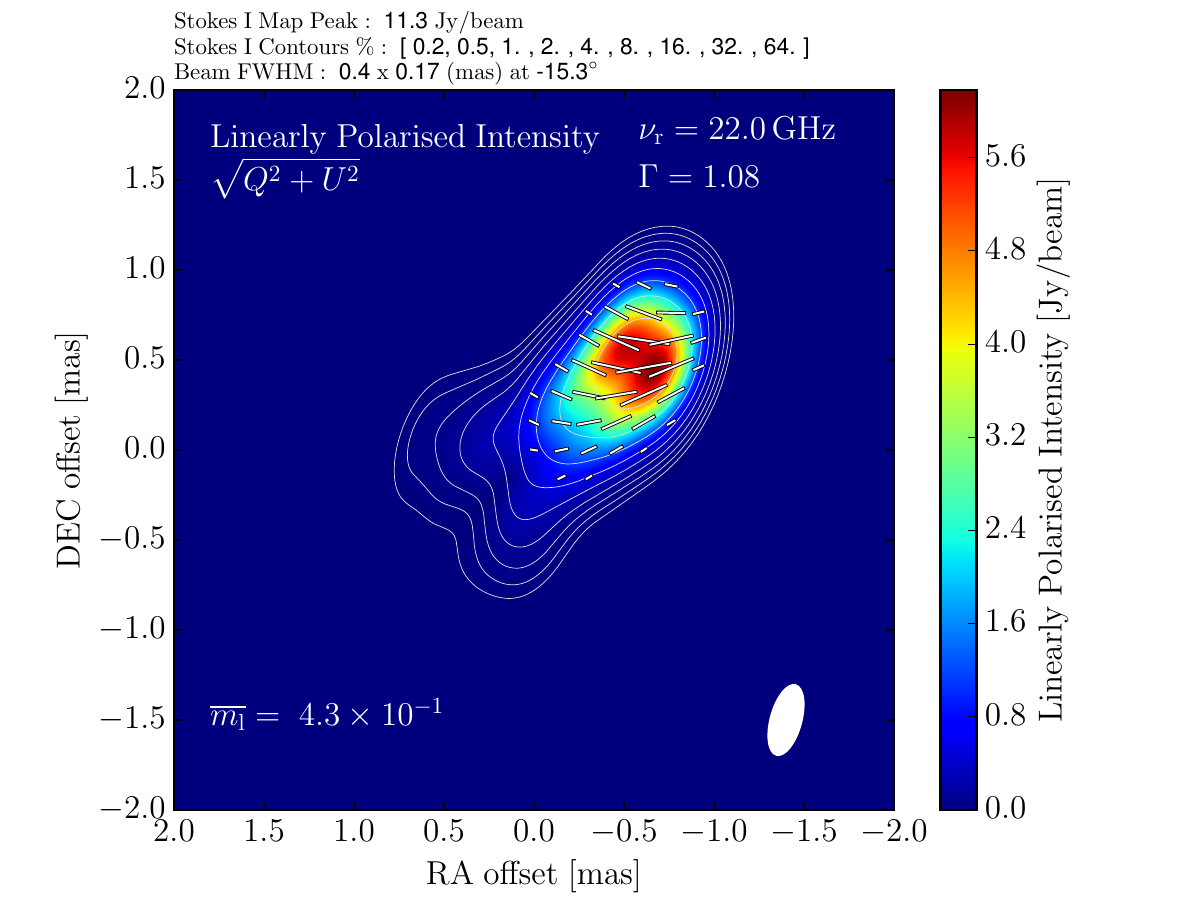} 
  \caption{
 Results from RMHD simulations.
 Ray-traced hybrid fluid-particle jet intensity maps are shown with a similar set-up to Fig.~\ref{fig:Pol} and altered magnetic field configuration.
 \textit{Left}: Poloidal magnetic field configuration results in a constant EVPA direction, dissimilar to the bimodality observed one.
 \textit{Right}: Helical magnetic field configuration results in the turn of the EVPAs, which differs from the observed configuration.
  }
     \label{fig:AppSimB}
\end{figure*}

Here we explore the effects that alternative parameters have on the outcome of the RMHD simulations.
Figure~\ref{fig:AppSimB} showcases the influence of the magnetic field configuration on the EVPAs and total intensity, for a fixed $\Gamma\sim1$.
In the left panel the poloidal magnetic field configuration produces an EVPA pattern that is at a constant angle to the bulk jet flow, and therefore disagreeing with the observations.
Furthermore, the right panel illustrates the effects of a helical magnetic field configuration on the jet flow.
Here the EVPAs exhibit some rotation, but again it does not match the observed pattern.
The Stokes $I$ contours exhibit some bifurcation, but only downstream and not reaching all the way into the compact core.
In summary, the toroidal magnetic field configuration (see Fig.~\ref{fig:Pol}) reproduces the observed linearly polarised structure more faithfully than the other two configurations that we investigated.

\begin{figure*}
\centering
\includegraphics[trim={0 0 0 0}, clip, scale=0.45]{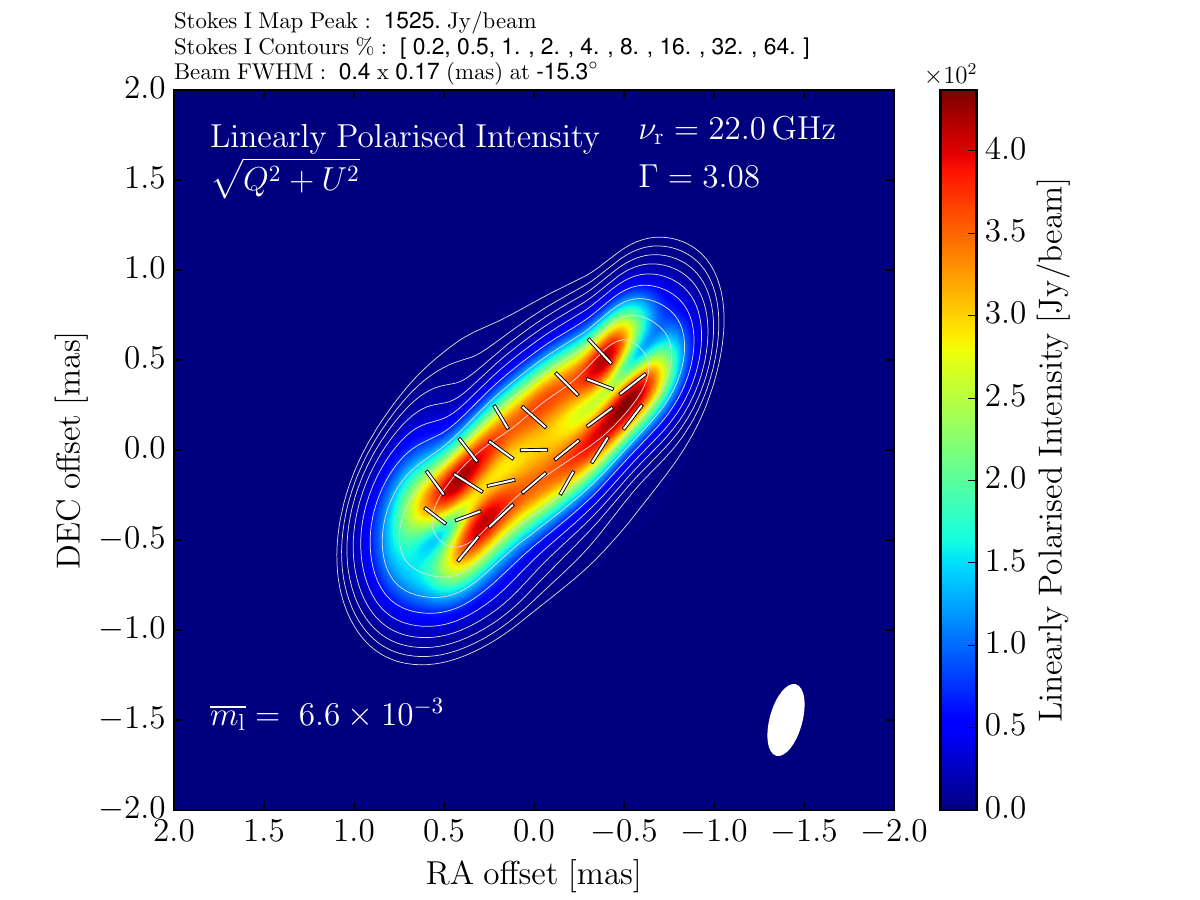} 
\includegraphics[trim={0 0 0 0}, clip, scale=0.45]{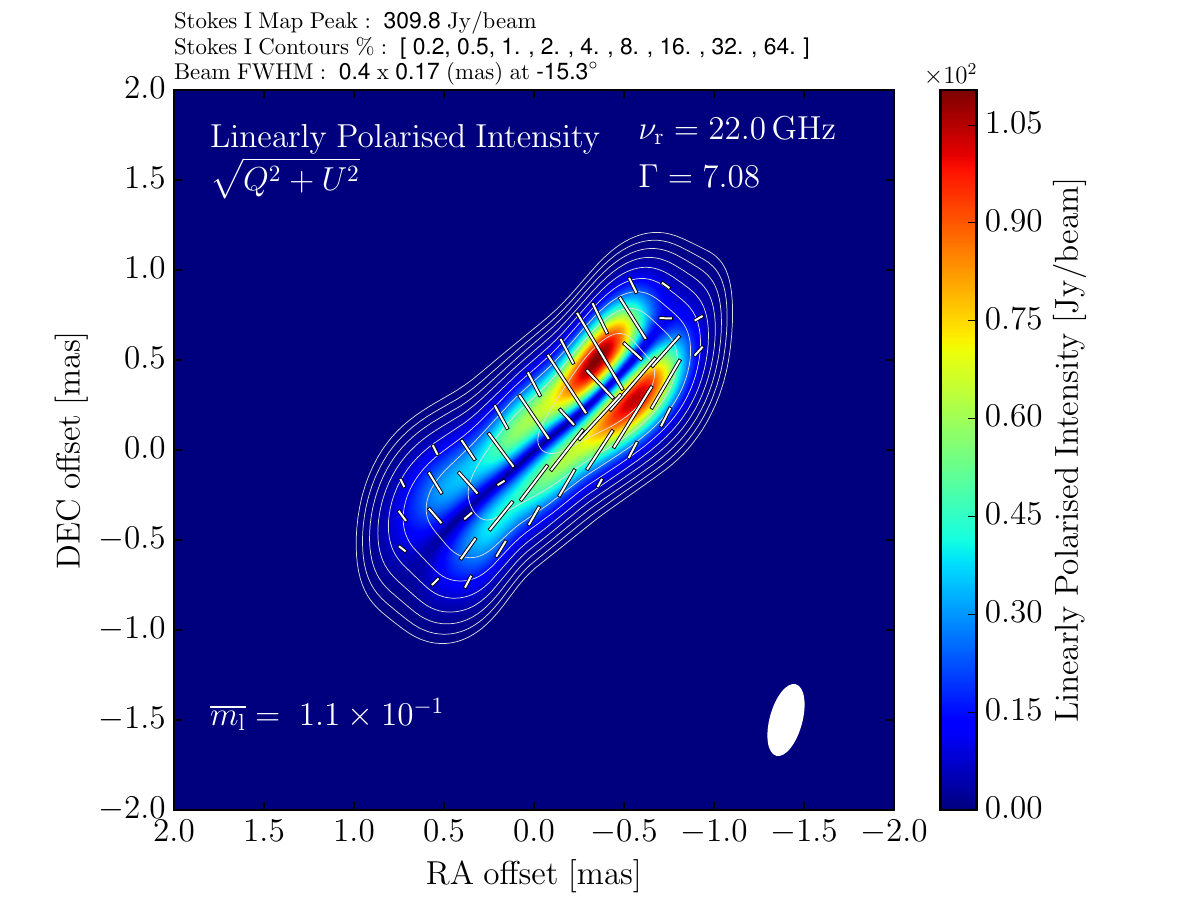} 
  \caption{
  Results from RMHD simulations.
  Ray-traced hybrid fluid-particle jet intensity maps are shown with a similar set-up to Fig.~\ref{fig:Pol} and altered $\Gamma$. 
  Both models carry an underlying toroidal magnetic field.
  The double rail structure is detected in both instances.
  The EVPA pattern remains similar to the $\Gamma=1.08$ case.
  \textit{Left}: For $\Gamma=3.08$ the extent of the core structure surpasses the observed structure.
  \textit{Right}: For $\Gamma=7.08$ the core structure stretches to a length almost double that of the observed structure.
  }
     \label{fig:AppSimGamma}
\end{figure*}

As a next step, we then constrained the magnetic field configuration to be toroidal and altered $\Gamma$.
In Fig.~\ref{fig:AppSimGamma} we present the effect of increasing $\Gamma$ on the simulated linearly polarised signatures.
The left panel corresponds to $\Gamma = 3.08$ and the right panel to $\Gamma = 7.08$. 
We find that in this case the EVPA patterns retain their bimodality.
The increase in $\Gamma$ also resulted in an apparent elongation of the limb-brightened structure of the core, due to Doppler boosting, overall exceeding the extent of the sub-mas region of our observations.
Therefore, our RMHD simulations and the observational evidence are both suggestive of only mildly relativistic velocities present in the sub-mas region of \C.

\end{appendix}

\end{document}